\documentstyle[prb,aps,eqsecnum,multicol,epsfig,color]{revtex}

\newcommand{\bleq}{\ifpreprintsty
                   \else
                   \end{multicols}\vspace*{-3.5ex}{\tiny
                   \noindent\begin{tabular}[t]{c|}
                   \parbox{0.493\hsize}{~} \\ \hline \end{tabular}}
                   \fi} 
\newcommand{\eleq}{\ifpreprintsty
                  \else
                   {\tiny\hspace*{\fill}\begin{tabular}[t]{|c}\hline
                    \parbox{0.49\hsize}{~} \\
                    \end{tabular}}\vspace*{-2.5ex}\begin{multicols}{2}
                    \fi}
\newcommand{\bcols}{\ifpreprintsty\else\begin{multicols}{2}\fi}
\newcommand{\ecols}{\ifpreprintsty\else\end{multicols}\fi}

\begin{document}
\draft

\title{Spin-density-wave instabilities in the organic conductor
  (TMTSF)$_2$ClO$_4$: Role of anion ordering. } 
\author{K. Sengupta$^{(1)}$ and N. Dupuis$^{(2)}$  }
\address{(1) Department of Physics, University of Maryland, College Park, MD
20742-4111, USA \\
(2) Laboratoire de Physique des Solides, Associ\'e au CNRS,
  Universit\'e Paris-Sud, 91405 Orsay, France }
\date{August 27, 2001}
\maketitle

\begin{abstract} 
We study the spin-density-wave (SDW) instabilities in the
quasi-one-dimensional conductor (TMTSF)$_2$ClO$_4$. The orientational
order of the anions ClO$_4$ doubles the unit cell and leads to the
presence of two electronic bands at the Fermi level. From
the Ginzburg-Landau expansion of the free energy, we
determine the low-temperature phase diagram as a function of the
strength of the Coulomb potential due to the anions. Upon increasing the anion
potential, we first find a  SDW phase corresponding to an
interband pairing. This SDW phase is rapidly suppressed, the metallic
phase being then stable down to zero temperature. The SDW instability
is restored when the anion potential becomes of the order of the
interchain hopping amplitude. The metal-SDW
transition corresponds to an intraband pairing which leaves half of
the Fermi surface metallic. At lower temperature, a second
transition, corresponding to the other intraband pairing, takes place
and opens a gap on the whole Fermi surface. We discuss the consequence
of our results for the experimental phase diagram of
(TMTSF)$_2$ClO$_4$ at high magnetic field. 
\end{abstract}


\bcols 


\section{Introduction}

The organic conductors of the Bechgaard salt family (TMTSF)$_2$X
(where TMTSF stands for tetramethyltetraselenafulvalene and X=PF$_6$,
ClO$_4$...) exhibit a very rich phase diagram when
temperature, magnetic field or pressure are varied.\cite{Kang93}
 One of the most
remarkable phenomena is the existence of a series of spin-density-wave
(SDW) phases in presence of a moderate magnetic field of a few
Tesla.\cite{reviews} These phases are separated by first-order
transitions and exhibit a quantization of the Hall effect:
$\sigma_{xy}=-2Ne^2/h$ per layer of TMTSF molecules, where the integer
$N$ varies at each phase transition.

According to the so-called quantized-nesting model (QNM),
\cite{reviews,Gorkov84}  the
formation of the magnetic-field-induced spin-density-wave (FISDW)
phases results from an interplay between the nesting properties of the
quasi-one-dimensional (Q1D)  
Fermi surface and the quantization of the electronic orbits in
magnetic field. Although the QNM explains the quantization of the Hall
effect\cite{Poilblanc87,Yakovenko91} and most features of the phase
diagram, deviations from the 
theoretical predictions have been observed at high magnetic field
in the compound (TMTSF)$_2$ClO$_4$. \cite{McKernan95,Moser99,Chung00}
In the last FISDW phase, when the magnetic field exceeds 18 Tesla, 
the second-order metal-SDW transition which occurs at
$T_c\simeq 5.5$ K is followed by a SDW-SDW transition at $T_c'\simeq
3.5$ K. It is believed that the existence of this low-temperature SDW
phase is due to the orientational ordering of the
(non-centrosymmetric) anions ClO$_4$ which occurs at $T_{\rm AO}\simeq
24$ K in slowly cooled (relaxed) samples. \cite{Kang93,McKernan95} 
Nevertheless, to our knowledge,
there is no satisfying theoretical description \cite{note1} of the
phase diagram of (TMTSF)$_2$ClO$_4$ at high field in spite of recent
progress. \cite{Kishigi97,Hasegawa98,Miyazaki99} 

In this paper, we study the effect of anion ordering on the
SDW phase in the absence of a magnetic field. In relaxed samples, the
ground-state of (TMTSF)$_2$ClO$_4$ is superconducting in the absence
of a magnetic field. Therefore, our study of the SDW in presence of
anion ordering does not apply directly to the experimental
situation. Nevertheless, it is the first step towards the
understanding of the behavior of (TMTSF)$_2$ClO$_4$ at high field. Our
work shows that in previous analysis 
\cite{Kishigi97,Hasegawa98,Lebed89,Osada92,Gorkov95} (with or without
magnetic field), the anion ordering was not accounted for properly. 
Because of the anion potential, the SDW order parameter necessarily
has two Fourier components. As a result, the metal-SDW transition
temperature is determined by a {\it generalized} Stoner
criterion. This point has been systematically overlooked, and this
calls for a revision of previous works.  

\begin{figure}
\centerline{\psfig{file=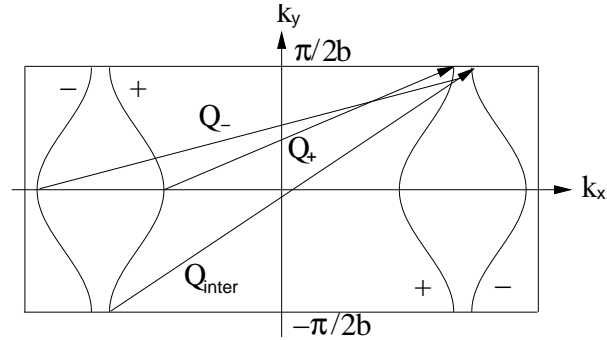,width=8cm,angle=0}}
\vspace{0.5cm}
\caption{Possible nesting vectors in presence of the anion
 potential. ${\bf Q}_{\rm inter}$ (${\bf Q}_\pm$) corresponds to
 interband (intraband) pairing. The magnitude of the band splitting at
 $k_y=\pm \pi/2b$ equals $2V$, where $V$ is the strength of the anion
 potential (see text). }
\label{fig1}
\end{figure}

The results obtained in this paper are based on a simple model, where
the anion ordering is assumed to create an electrostatic potential $V$
($-V$) on even (odd) chains. Moreover, the crystal structure is taken
to be orthorhombic, while the actual structure of the Bechgaard salts
is triclinic. In the spirit of the QNM, we expect such a simple model
(with a few unknown parameters) to correctly describe the physics of
(TMTSF)$_2$ClO$_4$. The anion potential $\pm V$ doubles the crystal
periodicity in the transverse direction. This leads to a reduced
Brillouin zone with two electronic bands crossing the Fermi level
(Fig.~\ref{fig1}) in qualitative agreement with the actual
Fermi surface of (TMTSF)$_2$ClO$_4$ as obtained from quantum chemistry
calculation.\cite{Peleven01} Instead of a
single (best) nesting vector, there are three possible nesting
vectors: ${\bf Q}_{\rm inter}$ (interband pairing), ${\bf Q}_+$ and
${\bf Q}_-$ (intraband pairing). The instability which does occur at low
temperature depends on the ratio $V/t_b$ between the anion potential
and the interchain hopping amplitude $t_b$. It has recently been shown,
both from quantum chemistry calculation \cite{Peleven01} and
experiments \cite{Yoshino97}, that $V$ can be of the order of $t_b$. 

In the next section, we calculate the electron-hole susceptibility
within the random-phase approximation (RPA) and thus obtain the
transition temperature between the metallic phase and the SDW
phase. The complete phase diagram is obtained from the Ginzburg-Landau
expansion of the free energy (Sec.~\ref{sec:gl}). For a weak anion
potential ($V\ll t_b$), we find that the SDW phase corresponds to the
interband pairing (${\bf Q}_{\rm inter}$). The transition temperature is
strongly suppressed by the anion potential and in general vanishes above a 
critical value of $V$. The SDW instability is restored when $V$
becomes of the order of the interchain hopping amplitude ($V\sim
t_b$). The metal-SDW 
transition corresponds to the intraband pairing ${\bf Q}_+$ (or ${\bf
Q}_-$). Half of the Fermi surface remains gapless, so that the SDW phase is
metallic. At lower temperature, a second SDW instability occurs at the
(intraband) nesting vector ${\bf Q}_-$ (or ${\bf  Q}_+$), thus
opening a gap on the whole Fermi surface. 

Some of our conclusions agree with the results of Kishigi {\it et al.}
\cite{Kishigi97,Hasegawa98,Miyazaki99}. In particular, these authors
have shown that the anion potential may stabilize SDW phases with wave
vectors ${\bf Q}_{\rm inter}$, ${\bf Q}_-$, ${\bf Q}_+$ (or ${\bf
Q}_-$ and ${\bf Q}_+$). \cite{note2}

\section{Instability of the metallic phase}

In Q1D materials, the Fermi surface consists of two slightly warped
open sheets. As a result, in the vicinity of the Fermi level, the electron
dispersion is well approximated as 
\begin{equation}
\epsilon_{\alpha}(k_x,k_y) = v_F(\alpha k_x -k_F) 
+ t_{\perp}(k_y b),
\label{disp}
\end{equation}
where $k_x$ and $k_y$ are the electron momenta along and across the
conducting chains, and $b$ is the interchain spacing. Here and in the
rest of the paper, we neglect the third direction ($z$ axis) which
does not play an important role for our purpose, and use natural units
$\hbar=k_B=c=1$. In Eq.~(\ref{disp}), the longitudinal electron
dispersion is linearized in $k_x$ in the vicinity of the two 1D Fermi
points $\pm k_F$, and $v_F=2at_a\sin(k_Fa)$ is the corresponding Fermi
velocity. $t_a$ is the transfer integral along the chain and $a$ the
lattice spacing. $\alpha = -,+$ corresponds to the left and the right
Fermi sheets. The transverse dispersion $t_{\perp}(k_y b)$ is given by
\begin{eqnarray}
t_{\perp}(k_y b)&=& t_{\perp}^{\rm odd}(k_y b)+ t_{\perp}^{\rm
even}(k_y b) \\
t_{\perp}^{\rm even}(k_y b)&=& -2t_{2b} \cos(2k_y b) \\
t_{\perp}^{\rm odd}(k_y b)&=&  -2t_b \cos(k_y b) -2t_{3b} \cos(3k_y b)
\label{transverse}
\end{eqnarray}
where $t_{nb}$ is the transfer integral for electron hopping to the
$n$th neighboring chain. $t_{\perp}^{\rm odd}$ and $t_{\perp}^{\rm
even}$ correspond to odd and even $n$, respectively.  For a simple
model with only nearest-neighbor hopping, {\it i.e.} when
$t_{2b}=t_{3b}=0$, the linearized dispersion (\ref{disp}) satisfies
the property $\epsilon_-({\bf k})=-\epsilon_+({\bf k}+{\bf Q}_0)$,
which corresponds to a perfect nesting of the Fermi surface at wave
vector ${\bf Q}_0=(2k_F,\pi/b)$. For $t_{2b}$, $t_{3b}\ne 0$, the
nesting becomes imperfect. Consequently, the nesting vector shifts to
${\bf Q}=(2k_F+\delta q_x,\pi/b+\delta q_y)$ and the SDW phase occurs
with lower transition temperature. Generally, $t_{3b}$ is neglected
since it is very small, $\it i.e.$, $t_{3b}/t_{2b} \ll 1$. However, as
we shall see, $t_{3b}$ plays an important role in the SDW phases with
intraband nesting vectors $Q_+$ and $Q_-$, and should therefore be
retained.

The anion potential in Q1D systems can be most simply modeled as
$V_{\rm anion} =V(-1)^n$, where $n$ is the chain index. In the presence
of the anion potential, the Hamiltonian of the system in 
the absence of electron-electron interaction can be written as  
\begin{eqnarray}
H_0 &=& \int d^2r \sum_{\alpha,\sigma} \hat \psi^{\dagger}_{\alpha
  \sigma} ({\bf r}) 
\Big[ v_F(\alpha \hat k_x -k_F) + t_{\perp} (\hat k_y b) \nonumber\\
&& + V(-1)^n \Big]\hat \psi_{\alpha \sigma} ({\bf r}) ,
\label{Hamiltonian}
\end{eqnarray}
where the $\hat\psi_{\alpha\sigma}$'s are fermionic operators for
right ($\alpha=+$) and left ($\alpha=-$) moving particles, and
$\sigma=\uparrow,\downarrow$ is the spin index. 
${\bf r}=(x,nb)$, and $\hat k_{x}$,$\hat k_{y}$ are
momentum operators along $x$ and $y$.

The Hamiltonian (\ref{Hamiltonian}) can be diagonalized to obtain the
eigenfunctions and energy eigenvalues:
\begin{eqnarray}
\psi^j_{\alpha,{\bf k}}({\bf r}) &=& \frac{1}{\sqrt A}
e^{i\left({\bf k+K}/2\right)\cdot{\bf r}} \sum_{p=\pm}
\gamma^{jp}_{k_y} e^{i p {\bf K}\cdot{\bf r}/2} ,  \
\label{eigenfn}\\
\epsilon_{\alpha,{\bf k}}^j &=& v_F(\alpha k_x -k_F) +
\epsilon_{k_y}^j ,
\label{energy}
\end{eqnarray}
where $j=\pm$, $A=L_x L_y$ is the area of the system,  ${\bf K}=
(0,\pi/b)$, and $k_y \in ]-\pi/2b,\pi/2b]$ reflecting period doubling along 
$y$ due to the presence of the anion potential. The transverse
momentum dependent part of the energy $\epsilon_{k_y}^j$ is given by 
\begin{eqnarray}
\epsilon_{k_y}^j &=& j \sqrt{ V^2 + \left[t_{\perp}^{\rm
 odd}(k_y b)\right]^2 } + t_{\perp}^{\rm even}(k_y b) .
\label{transen}
\end{eqnarray}
The presence of the anion potential splits the energy dispersions into two
bands with energies $\epsilon_{\alpha,{\bf k}}^+$ and
$\epsilon_{\alpha,{\bf k}}^-$ as shown in Fig.~\ref{fig1}. The factors
$\gamma^{jp}_{k_y}$ are given by 
\begin{eqnarray}
\gamma^{++}_{k_y}&=& \gamma^{--}_{k_y} = \frac{1}{\sqrt 2} \left( 1 -
\frac{t_{\perp}^{\rm odd}(k_y b)}{\sqrt{\left[ t_{\perp}^{\rm
odd}(k_y b) \right]^2+ V^2}} \right)^{1/2} , \nonumber\\
\gamma^{+-}_{k_y}&=& -\gamma^{-+}_{k_y} = \frac{1}{\sqrt 2} \left( 1 +
\frac{t_{\perp}^{\rm odd}(k_y b)}{\sqrt{\left[t_{\perp}^{\rm odd}(k_y
b) \right]^2+ V^2}} \right)^{1/2}.  
\label{gamma}
\end{eqnarray}
The amplitude of the wavefunction at ${\bf r}=(x,nb)$
depends on the factors $\gamma^{jp}_{k_y}$ and is given by
\begin{equation}
|\psi^j_{\alpha,{\bf k}}({\bf r})|=|\gamma^{j+}_{k_y}+(-1)^n
 \gamma^{j-}_{k_y}| .
\end{equation}
Electronic states in the $+$ ($-$) band have a
higher probability amplitude on even (odd) chains. This localization
becomes very important when $V\gtrsim \sqrt{2}t_b$ (assuming
$t_{3b}\ll t_{b}$, which is obviously the case for a realistic
dispersion law). In this regime,
even and odd chains tend to decouple, and the dispersion law reduces
to $\epsilon_{\alpha,{\bf k}}^j \simeq v_F(\alpha k_x-k_F)+jV
+t_\perp^{\rm even}(k_yb)$. 

Using Eqs.~(\ref{eigenfn}) and (\ref{energy}), we obtain the Green
functions in the absence of electron-electron interaction: 
\begin{eqnarray}
G_{\alpha\sigma}({\bf r},{\bf r'};\omega_n) &=& \sum_{j,{\bf k}} 
\frac{\psi^j_{\alpha,{\bf k}}({\bf
r})\psi^{j\,*}_{\alpha,{\bf k}}({\bf r}')}{i \omega_n - 
\epsilon_{\alpha,{\bf k}}^j },
\label{Green}
\end{eqnarray}
where $\omega_n$ is a fermionic Matsubara frequency.

\subsection{Bare susceptibilities}

Due to the presence of two electronic bands, there are three possible
spin-density-wave (SDW) instabilities with wave-vectors
${\bf Q}_{+}$, ${\bf Q}_{-}$, and ${\bf Q}_{\rm inter}$ as shown in
Fig.~\ref{fig1}. ${\bf Q}_{+}$ and ${\bf Q}_{-}$ are intraband nesting vectors
which satisfy $(Q_y)_{\pm}\approx \pi/2b$.
${\bf Q}_{\rm inter}$ is the inter-band nesting vector (with $(Q_y)_{\rm
inter} \approx \pi/b$). In the absence of the anion 
potential, the SDW instability occurs with ${\bf Q}={\bf Q}_{\rm
inter}$. To obtain a quantitative description of these possible
instabilities and the phase diagram for the system, we calculate the
susceptibility $\chi({\bf q},{\bf q}')$  in the particle-hole
channel within the RPA. First, let us consider the susceptibilities
in the absence of electron-electron interaction: 
\begin{eqnarray}
\chi_{\alpha \sigma}^0({\bf r},{\bf r'};\tau-\tau') &=& \langle 
{\mathcal T}_\tau \hat \Delta_{\alpha \sigma}({\bf r},\tau) \hat
\Delta_{\alpha \sigma}^{\dagger}({\bf r'},\tau')\rangle _0 ,
\label{baresus}
\end{eqnarray} 
where $\hat\Delta_{\alpha \sigma}({\bf r},\tau)= \hat \psi_{\bar
\alpha\bar \sigma}^{\dagger}({\bf r},\tau) \hat
\psi_{\alpha\sigma}({\bf r},\tau)$, $\tau$ is an imaginary time, and
${\mathcal T}_\tau$ is the time-ordering operator. The mean value in
Eq.~(\ref{baresus}) has to be taken with the Hamiltonian $H_0$. We use
the notation $\bar\alpha=-\alpha$, and
$\bar\sigma=\downarrow,\uparrow$ for $\sigma=\uparrow,\downarrow$. In
frequency domain, $\chi^0$ can be expressed in terms of the Green
function (\ref{Green}) as
\begin{equation}
\chi_{\alpha \sigma}^0({\bf r},{\bf r'};p_n) = - T \sum_{\omega_n} 
G_{\alpha \sigma}({\bf r},{\bf r'};\omega_n) G_{\bar \alpha \bar
\sigma} ({\bf r'},{\bf r};\omega_n - p_n),
\label{freqsus}
\end{equation}
where $p_n$ is a bosonic Matsubara frequency. 
For studying the instabilities of the metallic phase, it is sufficient to
compute the static susceptibilities $\chi_{\alpha \sigma}^0({\bf r},{\bf
r'};p_n=0) \equiv \chi_{\alpha \sigma}^0({\bf r},{\bf r'})$. Using
Eqs.~(\ref{eigenfn}), (\ref{energy}), (\ref{gamma}), and  
(\ref{Green}), we find  
\begin{eqnarray}
&& \chi_{\alpha \sigma}^0({\bf q},{\bf q})= \frac{1}{A} \sum_{\bf k} \Bigg{\{}
\left[\overline \chi_{\alpha \sigma}^{++}({\bf k},{\bf q})+ \overline
\chi_{\alpha \sigma}^{--}({\bf k},{\bf q}) \right] \nonumber\\
&& \times \left(
\gamma^{++}_{k_y} \gamma^{++}_{k_y-q_y} +
\gamma^{+-}_{k_y}\gamma^{+-}_{k_y-q_y }\right)^2 + \left[\overline
\chi_{\alpha \sigma}^{+-}({\bf k},{\bf q})+ \overline 
\chi_{\alpha \sigma}^{-+}({\bf k},{\bf q}) \right] \nonumber\\
&& \times \left(
\gamma^{++}_{k_y} \gamma^{+-}_{k_y-q_y} -
\gamma^{++}_{k_y-q_y}\gamma^{+-}_{k_y}\right)^2 \Bigg{\}} ,
\label{diasus0} \\
&&\chi_{\alpha \sigma}^0({\bf q},{\bf q+K})= \frac{1}{A} \sum_{\bf
k} \Bigg{\{} 
\left[\overline \chi_{\alpha \sigma}^{++}({\bf k},{\bf q})- \overline
\chi_{\alpha \sigma}^{--}({\bf k},{\bf q}) \right] \nonumber\\
&& \times \left(
\gamma^{++}_{k_y} \gamma^{+-}_{k_y} +
\gamma^{++}_{k_y-q_y}\gamma^{+-}_{k_y-q_y }\right) + \left[\overline
\chi_{\alpha \sigma}^{+-}({\bf k},{\bf q})- \overline 
\chi_{\alpha \sigma}^{-+}({\bf k},{\bf q}) \right] \nonumber\\
&& \times \left(
\gamma^{++}_{k_y} \gamma^{+-}_{k_y} -
\gamma^{++}_{k_y-q_y}\gamma^{+-}_{k_y-q_y}\right) \Bigg{\}},
\label{offdiasus0}
\end{eqnarray}
where 
\begin{eqnarray}
&&\overline \chi_{\alpha \sigma}^{jj'}({\bf k},{\bf q}) = 
-T \sum_{\omega_n} \left(i\omega_n - \epsilon_{\alpha,{\bf k}}^{j}
\right)^{-1} \left(i\omega_n - \epsilon_{\bar \alpha,{\bf k-q}}^{j'}
\right)^{-1} . \nonumber\\
\label{chi}
\end{eqnarray}
Notice that due to the
presence of the anion potential, the static susceptibilities have a 
non-zero off-diagonal component $\chi^0({\bf q},{\bf q +K})$. The sum
over $k_x$ in Eqs.~(\ref{diasus0}) and (\ref{offdiasus0}) can be
analytically calculated using 
\begin{eqnarray}
&& \frac{1}{bL_x} \sum_{k_x} \overline \chi_{\alpha \sigma}^{jj'}({\bf
k},{\bf q}) = \frac{N(0)}{2} \left[ \log \left(\frac{2 \gamma E_0}{\pi
T}\right) 
+ \Psi \left(\frac{1}{2}\right) \right. \nonumber\\
&& \left. - {\rm Re} \, \Psi \left(\frac{1}{2}- \frac{ v_F
(\alpha q_x -2 k_F) + \epsilon_{k_y}^{j} + \epsilon_{k_y-q_y}^{j'} }{4\pi i
T}\right) \right],
\label{chieval}
\end{eqnarray} 
but the $k_y$ sum needs to be evaluated numerically. In
Eq.~(\ref{chieval}), $N(0)=1/\pi v_Fb$ is the density of states per
spin at Fermi energy, $E_0 \sim t_a$ an ultraviolet cutoff energy,
$\gamma \simeq 1.783$ the exponential of the Euler constant, $\Psi$
the digamma 
function, and ${\rm Re}\, \Psi$ means real part of $\Psi$. Note that when  
$V=0$, $\chi^0_{\alpha \sigma} ({\bf k},{\bf k+q}) =0$ and the
susceptibility becomes diagonal in momentum space. 

\begin{figure}
\epsffile[85 193 369 418]{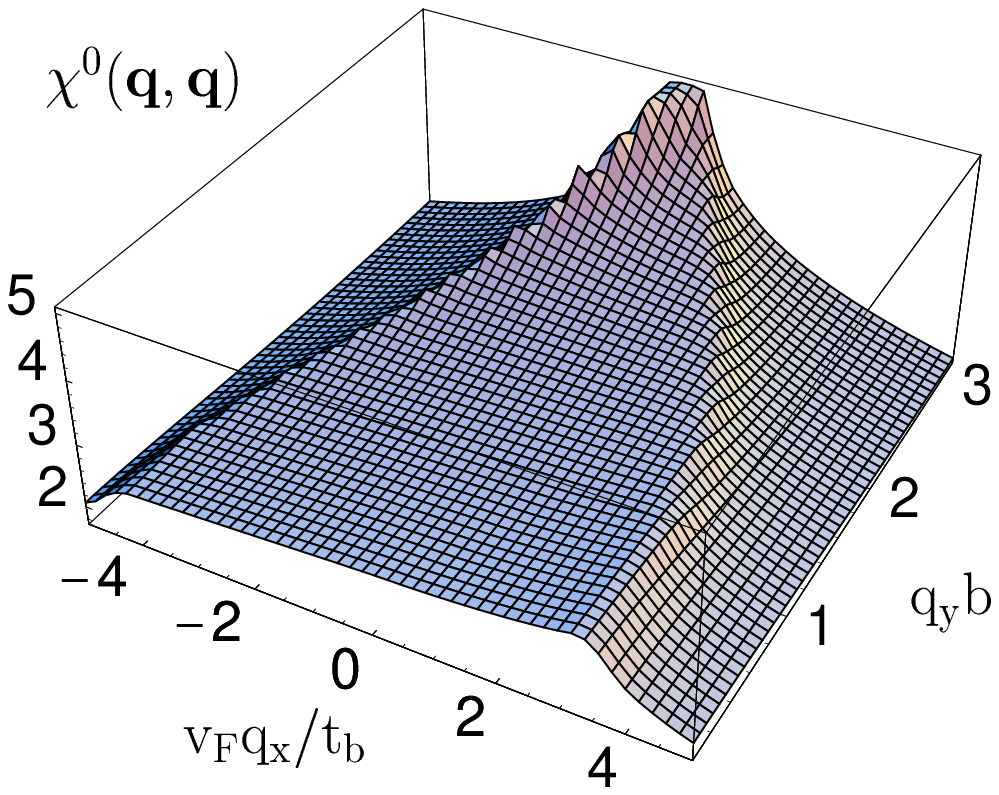}
\caption{Susceptibility $\chi^0({\bf q},{\bf q})$ in the absence of
anion ordering ($V=0$). Here $\chi^0\equiv
\chi^0_{+\uparrow}=\chi^0_{+\downarrow}$, and $q_x$ is measured from
$2k_F$.The maximum is located at the interband nesting vector ${\bf
Q}_{\rm inter}$ with $(Q_{\rm inter})_y \approx \pi/b$. The
susceptibilities shown in Figs.~\ref{fig:V0d}-\ref{fig:V1d} are
obtained for $T=0.02t_b$, $t_{2b}=0.1 t_b$ and $t_{3b}=0.02t_b$.}
\label{fig:V0d}
\end{figure}

\begin{figure}
\epsfxsize 8cm
\epsffile[71 194 369 417]{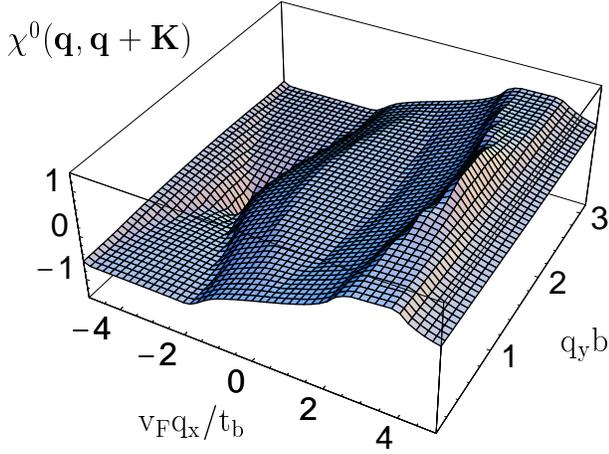}
\caption{
Off-diagonal component  $\chi^0({\bf q},{\bf q}+{\bf K})$ of the
susceptibility for $V/t_b=1$. The notations are the same as in
Fig.~\ref{fig:V0d}.}
\label{fig:V1od}
\end{figure}

\begin{figure}
\epsfxsize 8cm
\epsffile[85 197 369 414]{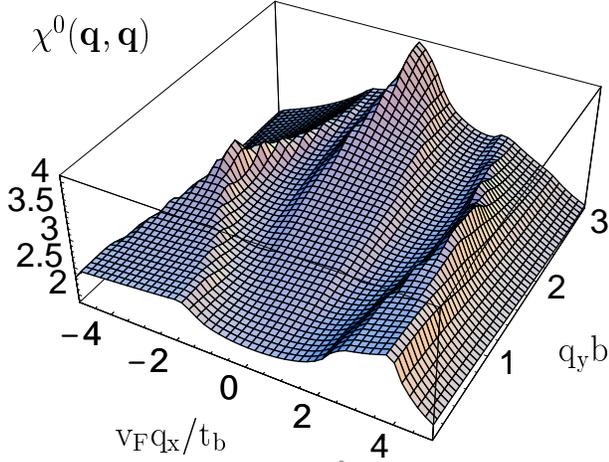}
\caption{Diagonal component  $\chi^0({\bf q},{\bf q})$ of the
susceptibility for $V/t_b=1$. Besides the maximum at ${\bf Q}_{\rm
inter}$, there are two peaks at $q_y=\pi/2b$ corresponding to the
intraband nesting vectors ${\bf Q}_+$ and ${\bf Q}_-$. The notations
are the same as in Fig.~\ref{fig:V0d}.}
\label{fig:V1d}
\end{figure}

To find the effect of the anion potential $V$ on the bare
susceptibilities, we plot $\chi^0({\bf q},{\bf q})$ and $\chi^0({\bf
q},{\bf q+K})$ for different values of $V/t_b$ in
Figs.~\ref{fig:V0d}-\ref{fig:V1d}. Here $\chi^0\equiv 
\chi^0_{+\uparrow}=\chi^0_{+\downarrow}$, and ${\bf q}$ is chosen
such that $q_x \sim 2k_F$. For $V/t_b=0$, $\chi^0$ is
diagonal in momentum space and the peak of $\chi^0({\bf q},{\bf q})$
is located at ${\bf Q}_{\rm inter}$ with $(Q_{\rm inter})_y \approx
\pi/b$ (Fig.~\ref{fig:V0d}).  The position of the peak moves away from
$\pi/b$ when deviations from perfect nesting due to $t_{2b}$ and
$t_{3b}$ become important. As we increase $V/t_b$, $\chi^0({\bf
q},{\bf q+K})$ becomes non-zero and develops peaks at $q_y \approx
\pm\pi/2b$ as shown, for $V/t_b=1$, in Fig.~\ref{fig:V1od}. The
maximum of $\chi^0({\bf q},{\bf q})$ ({\it i.e.} $\chi^0({\bf Q}_{\rm
inter},{\bf Q}_{\rm inter}$)) reduces in height and two additional
peaks develop at ${\bf Q}_+$ and ${\bf Q}_-$ (with
$(Q_\pm)_y=\pi/2b$). The development of these additional peaks in
$\chi^0({\bf q},{\bf q})$ can be seen by comparing Figs.~\ref{fig:V0d}
and \ref{fig:V1d}. When $V$ is strong enough, the maximum of $\chi^0$
moves from ${\bf Q}_{\rm inter}$ to ${\bf Q}_+$ or ${\bf Q}_-$. We
therefore expect the SDW wave-vector to shift from ${\bf Q}_{\rm
inter}$ to ${\bf Q}_{+}$ and/or ${\bf Q}_-$ as $V$ exceeds a critical
value (assuming that the anion potential $V$ does not suppress the SDW
instability). Our results for the diagonal susceptibility $\chi^0({\bf
q},{\bf q})$ are similar to those of Ref.\onlinecite{Miyazaki99}.

\subsection {RPA calculation}
\label{subsec:rpa}

To find the critical value of the anion potential and to obtain
the phase diagram, we now compute the susceptibilities for the
interacting system within RPA. We model the interaction
using the g-ology model keeping only $g_2$ to be non-zero:
\cite{comment3}
\begin{eqnarray}
H_{\rm int} &=& \frac{g_2}{2} \sum_{\alpha,\sigma,\sigma'}
\int d^2 r \hat \psi^{\dagger}_{\alpha \sigma} ({\bf r}) \hat
\psi^{\dagger}_{\bar \alpha \sigma'} ({\bf r}) \hat
\psi_{\bar \alpha
\sigma'} ({\bf r}) \hat \psi_{\alpha \sigma} ({\bf r}) .
\label{intham}
\end{eqnarray}

\begin{figure}
\centerline{\psfig{file=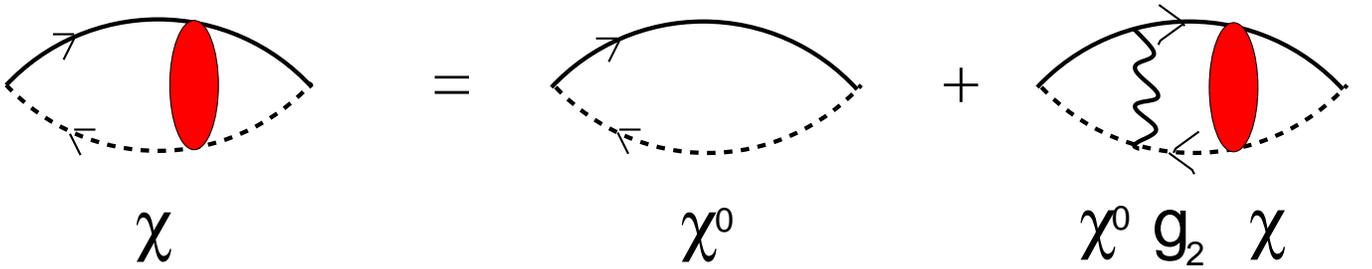,width=\linewidth,angle=0}}
\caption{Feynman diagrams  for $\chi$ within RPA.}
\label{fig4}
\end{figure}

Using the interaction Hamiltonian given by Eq.~(\ref{intham}), we obtain
the susceptibilities by summing the RPA diagrams shown in
Fig.~{\ref{fig4}}:
\begin{eqnarray}
\chi_{\alpha \sigma} ({\bf q},{\bf q'}) &=& \chi^0_{\alpha \sigma}
({\bf q},{\bf q'}) + g_2 \sum_{\bf q''} \chi^0_{\alpha \sigma}
({\bf q},{\bf q''}) \chi_{\alpha \sigma}
({\bf q''},{\bf q'}). \nonumber\\
\label{suseq}
\end{eqnarray}
Substituting Eqs.~(\ref{diasus0}) and (\ref{offdiasus0}) in
Eq.~(\ref{suseq}), we find that $\chi_{\alpha \sigma}  
({\bf q},{\bf q'})$ is non-zero only when ${\bf q'}={\bf q}$ or ${\bf
q+K}$ and is given by 
\begin{eqnarray}
&& \chi_{\alpha \sigma} ({\bf q},{\bf q}) = \Big \{
\chi^0_{\alpha \sigma} ({\bf q},{\bf q}) [1-g_2 \chi_{\alpha \sigma}^0
({\bf q+K},{\bf q+K})] \nonumber\\
&& \qquad \qquad + g_2 [\chi_{\alpha \sigma}^0 ({\bf q},{\bf
q+K})]^2\Big \} /D ,
\label{ds}\\
&& \chi_{\alpha \sigma} ({\bf q},{\bf q+K}) =
\chi^0_{\alpha \sigma} ({\bf q},{\bf q+K}) / D ,
\label{ods}\\
&& D = [1-g_2 \chi_{\alpha \sigma}^0
({\bf q},{\bf q})][1-g_2 \chi_{\alpha \sigma}^0
({\bf q+K},{\bf q+K})]\nonumber\\
&& \qquad \qquad -g_2^2 [\chi_{\alpha \sigma}^0 ({\bf q},{\bf q+K})]^2 .
\label{deno}
\end{eqnarray}
A SDW instability occurs when the susceptibilities $\chi^0_{\alpha
\sigma} ({\bf q},{\bf q})$ and $\chi^0_{\alpha \sigma} ({\bf q},{\bf
q+K})$ diverge, {\it i.e.} when the denominator $D$ in Eqs.~(\ref {ds})
and (\ref{ods}) vanishes. This yields a generalized
Stoner criterion for a SDW instability at wave vector ${\bf Q}$:
\begin{eqnarray}
&& [1-g_2 \chi^0
({\bf Q},{\bf Q})][1-g_2 \chi^0({\bf Q}+{\bf K},{\bf Q}+{\bf K})] \nonumber\\ 
&& -g_2^2 [\chi^0 ({\bf Q},{\bf Q}+{\bf K})]^2 = 0 ,
\label{stoner}
\end{eqnarray}
where ${\bf Q}$ is chosen such that $Q_x\sim 2k_F$. This equation is
the same for ${\bf Q}$ and ${\bf Q}+{\bf K}$ (reflecting the fact that
the SDW contains Fourier components at ${\bf Q}$ and ${\bf Q}+{\bf
K}$). We choose to label the SDW phase by the wave vector ${\bf Q}$
such that $\chi^0 ({\bf Q},{\bf Q})> \chi^0({\bf Q}+{\bf
  K},{\bf Q}+{\bf K})$. Note that the
off-diagonal component of the susceptibility, {\it i.e.} $\chi^0 ({\bf
  Q},{\bf Q}+{\bf K})$, has been systematically overlooked in previous
works on the effect of anion ordering in (TMTSF)$_2$ClO$_4$. As a
result, the simple (but wrong) criterion $1-g_2\chi^0({\bf Q},{\bf
  Q})=0$ has been used to determine the SDW transition temperature
instead of the generalized Stoner criterion given in
Eq.~(\ref{stoner}). 

Eq.~(\ref{stoner}) has to be numerically solved to obtain the
transition temperature $T_c(V)$ as a function of the anion
potential $V$. The qualitative nature of the phase diagram, however, can
be understood qualitatively from Eq.~(\ref{stoner}) without numerical
computation. For $V=0$, the off-diagonal susceptibility $\chi^0 ({\bf
  Q},{\bf Q+K})=0$ and Eq.\ (\ref{stoner}) reduces to 
$1-g_2 \chi^0 ({\bf Q},{\bf Q})=0$. The SDW instability occurs with
the nesting wave-vector ${\bf Q}_{\rm inter}$ corresponding to the
maximum of the bare susceptibility $\chi^0$. When $V$ increases, the
wave functions become more and more localized on even or odd
chains. Since ${\bf Q}_{\rm inter}$ corresponds to pairing of electron
and hole on chains of opposite parity, $T_c$ drops, until at some
critical value of the anion potential, $V_{c1}$, there is no SDW
instability with ${\bf Q}={\bf Q}_{\rm inter}$. [Note that the anion
potential does not affect the nesting at ${\bf Q}_{\rm inter}$, which
is limited by $t_{2b}$ in our model.]   
Simultaneously, the pairing with ${\bf
Q}={\bf Q}_{+}$ and/or ${\bf Q}_{-}$ becomes increasingly favorable,
due to both the wave function localization and an improved nesting, and
eventually at a value of the anion potential, $V_{c2}\sim t_b$, the SDW
phase with ${\bf Q}={\bf Q}_{+}$ and/or ${\bf Q}_{-}$ wins over the
metallic phase.
Indeed, when $V\gg t_b$, the even and the odd chains tend to decouple. The
dispersion law of the two electronic bands ($+$ and $-$) becomes 
$\epsilon^j_{\alpha,{\bf k}} \simeq v_F(\alpha k_x-k_F)+jV
+t_\perp^{\rm even}(k_yb)$, and $T_c(V) \to  T_c^{(0)}
= \left(2\gamma E_0 /\pi\right) \exp(-2/N(0)g_2)$, where $T_c^{(0)}$
is the transition temperature for a system with perfect nesting. 
Higher-order harmonics in the transverse dispersion law, such as
$t_{4b}$ (not considered in this paper), would introduce deviations
from perfect nesting that survive 
when $V\to\infty$. They are however expected to be very small and can
be safely discarded. 
If $V_{c2}>V_{c1}$, there will be a metallic region in the
phase diagram with no SDW instability for $V_{c1}<V<V_{c2}$;
otherwise, there will be a 
transition from the SDW phase with ${\bf Q}={\bf Q}_{\rm inter}$ to
the SDW phase with ${\bf Q}={\bf Q}_{+}$ and/or ${\bf Q}_{-}$.

\begin{figure}
\centerline{\psfig{file=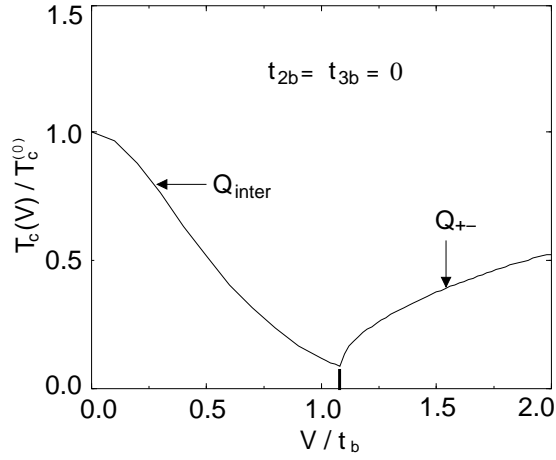,width=7.25cm,angle=0}}
\vspace{0.25cm}
\caption{Phase diagram with $t_{2b}=t_{3b}=0$. The thick vertical line
is a guide to the eye for estimating the critical potential at which
the transition from ${\bf Q}_{\rm inter}$ to ${\bf Q}_{\pm}$ takes
place. For $V>V_c \simeq 1.1 t_b$, there is coexistence of two SDW's
(with wave vectors ${\bf Q}_+$ and ${\bf Q}_-$) below the transition
line (see Sec.~\ref{subsec:pd}). 
}
\label{fig5}
\end{figure}

Our numerical calculations show that both the scenarios are possible. 
The phase diagram with $t_{2b}=t_{3b}=0$ is shown in
Fig.~\ref{fig5}. In this case, there is a transition between SDW
phases with ${\bf Q}= {\bf Q}_{\rm inter}$ and ${\bf Q}= {\bf
Q}_{\pm}$ without any intermediate metallic phase. Moreover, the SDW's
with ${\bf Q}= {\bf Q}_{\pm}$ have the same transition temperature. 
The transition between the SDW phases with ${\bf Q}_{\rm inter}$ and
${\bf Q}_{\pm}$ occurs at $V= V_c \approx 1.1 t_b$, as indicated by
the thick vertical line in Fig.~\ref{fig5}.

\begin{figure}
\centerline{\psfig{file=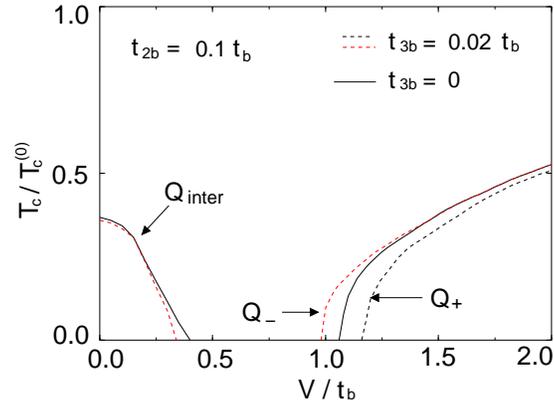,width=7.25cm,angle=0}}
\vspace{0.25cm}
\caption{Phase diagram with non-zero $t_{2b}$ and $t_{3b}$ showing the
intermediate metallic region. The solid line indicates the transition
temperature when $t_{3b}=0$. Below $T_c^{\rm coex} \simeq T_c^+$, the
two intraband SDW's coexist (see Sec.~\ref{subsec:pd}). 
}
\label{fig6}
\end{figure}

The phase diagram with more realistic parameters, shown in Fig.~\ref{fig6},
is quite different. With $t_{2b}=0.1t_b$ and $t_{3b}=0$, the SDW phase
with ${\bf Q}_{\rm inter}$ persists up to $V_{c1} \approx 0.4 t_{b}$. For
$V>V_{c1}$, we find a metallic phase till $V=V_{c2} \approx
t_{b}$. For $V>V_{c2}$, the SDW phase with ${\bf Q}_\pm$ is
stabilized. When $t_{2b}$ is strong enough, the SDW at ${\bf Q}_{\rm
inter}$ is suppressed ({\it i.e.} $V_{c1}=0$), but the transition
temperature of the intraband SDW's is not affected. In this case, the
metallic phase is stable at $T=0$ up to $V_{c2}$ where the SDW with
wave vector ${\bf Q}_+$ or ${\bf Q}_-$ sets in. 

The degeneracy between ${\bf Q}_+$ and ${\bf Q}_-$ is
lifted by the presence of a non-zero $t_{3b}$
(Fig.~\ref{fig6}). For $t_{3b}>0$, the SDW with wave vector ${\bf
  Q}_-$ has a higher transition temperature, {\it i.e.}
$T_c^->T_c^+$. The opposite is true when $t_{3b}<0$. It is clear that
when $V\gg t_b$, the two SDW's will coexist at low temperature. In
this regime, even and odd chains are essentially decoupled, and the
two (intraband) instabilities take place almost independently at transition
temperatures $T_c^+$ and $T_c^-$. This leads to a phase with
two coexisting SDW's below $T_c^{\rm coex}\simeq T_c^+<T_c^-$ (assuming
$t_{3b}>0$). 

\begin{figure}
\centerline{\psfig{file=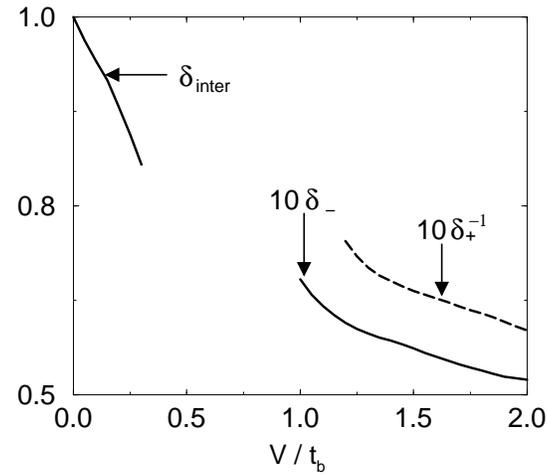,width=8.1cm,angle=0}}
\caption{
Ratio $\delta=(1+\tan(\theta))/(1-\tan(\theta))$ of the SDW amplitudes
on even and odd chains (see Sec.~\ref{subsec:sm}). The figure shows
$\delta$ if $\delta<1$ and $\delta^{-1}$ if $\delta>1$. The parameters
of the plot are the same as in Fig.~\ref{fig6} (with
$t_{3b}=0.02t_b$). }  
\label{fig:theta}
\end{figure}

Fig.\ \ref{fig:theta} shows the ratio 
$\delta=(1+\tan(\theta))/(1-\tan(\theta))$ of the SDW amplitudes
on even and odd chains (see Sec.~\ref{subsec:sm}).

\section{Ginzburg-Landau expansion}
\label{sec:gl}

In the preceding section, we have shown that two SDW instabilities (at
wave vectors ${\bf Q}_+$ and ${\bf Q}_-$) may occur when $V$ is strong
enough. In order to study in more detail the possibility of coexisting SDW's, 
we derive the Ginzburg-Landau expansion of the free energy and then
deduce the phase diagram.  

In the presence of two SDW's, the order parameter
$\Delta_{\alpha\sigma}({\bf r})=\langle \hat
\Delta_{\alpha\sigma}({\bf r}) \rangle $ takes the general form
\begin{equation}
\Delta_{\alpha\sigma}({\bf r})= \sum_{j=\pm,p=\pm}
\Delta_{\alpha\sigma}^{jp} e^{i\alpha({\bf Q}_j+(p-1)\frac{\bf
    K}{2})\cdot {\bf r} } .
\label{op1}
\end{equation} 
Note that the SDW with wave vector ${\bf Q}_j$ corresponds to a spin
modulation with Fourier components ${\bf Q}_j$ and ${\bf Q}_j+{\bf K}$. 
The order parameter (\ref{op1}), which corresponds to electron-hole
pairs with opposite spins, assumes the SDW's to be polarized in the
$(x,y)$ plane, {\it i.e.} $\langle S_z({\bf r})\rangle=0$. 
The spin modulation $\langle {\bf S}({\bf r})\rangle $ is discussed in
more detail below (Sec.~\ref{subsec:sm}). 
The relation $\Delta_{\bar\alpha\bar\sigma}({\bf r})=
\Delta_{\alpha\sigma}^*({\bf r})$ implies that the complex order
parameters $\Delta_{\alpha\sigma}^{jp}$ satisfy
$\Delta_{\bar\alpha\bar\sigma}^{jp}=
{\Delta_{\alpha\sigma}^{jp}}^*$. 

Up to quartic
order in the order parameter, the free energy (per unit surface) is given by
\bleq
\begin{eqnarray}
F &=& g_2 \sum_\alpha \int \frac{d^2r}{A}
|\Delta_\alpha({\bf r})|^2 + g_2^2 \frac{T}{A}
\sum_{\alpha,\omega} \int d^2r_1 d^2r_2 \Delta_\alpha^*({\bf r}_1)
\Delta_\alpha({\bf r}_2) G_{\alpha\uparrow}({\bf r}_1,{\bf r}_2,\omega)
G_{\bar\alpha\downarrow}({\bf r}_2,{\bf r}_1,\omega)  
\nonumber \\ && 
+ g_2^4 \frac{T}{2A} \sum_{\alpha,\omega}
\int d^2r_1d^2r_2d^2r_3d^2r_4 
\Delta_\alpha^*({\bf r}_1) \Delta_\alpha({\bf r}_2)
\Delta_\alpha^*({\bf r}_3) \Delta_\alpha({\bf r}_4)
\nonumber \\ && \times
G_{\alpha\uparrow}({\bf r}_1,{\bf r}_2,\omega)
G_{\bar\alpha\downarrow}({\bf r}_2,{\bf r}_3,\omega)
G_{\alpha\uparrow}({\bf r}_3,{\bf r}_4,\omega)
G_{\bar\alpha\downarrow}({\bf r}_4,{\bf r}_1,\omega) ,
\label{energy1}
\end{eqnarray}
where $\Delta_\alpha({\bf r})\equiv \Delta_{\alpha\uparrow}({\bf r})$.

\subsection{Quadratic contribution $F_2$}
\label{subsec:F2}

Let us first consider the quadratic contribution $F_2$ to the free
energy. One easily obtains
\begin{equation}
F_2 = g_2 \sum_{\alpha,j} ({\Delta_\alpha^{j+}}^*, {\Delta_\alpha^{j-}}^*) 
\left (
\begin{array}{lr}
 1-g_2 \chi^0({\bf Q}_j,{\bf Q}_j)  &  -g_2  \chi^0({\bf Q}_j,{\bf
  Q}_j+{\bf K})  \\ -g_2  \chi^0({\bf Q}_j+{\bf K},{\bf Q}_j)
 & 1-g_2 \chi^0({\bf Q}_j+{\bf K},{\bf Q}_j+{\bf K})
\end{array}
\right )
\left (
\begin{array}{l}
\Delta_\alpha^{j+} \\  \Delta_\alpha^{j-}
\end{array}
\right ) .
\end{equation}
Introducing the new order parameters $u_{j\alpha},v_{j\alpha}$ defined
by 
\begin{equation}
\left(
\begin{array}{l}
\Delta_\alpha^{j+} \\  \Delta_\alpha^{j-}
\end{array}
\right)
= 
\left( 
\begin{array}{lr}
\cos(\theta_j) & -\sin(\theta_j) \\
\sin(\theta_j) & \cos(\theta_j) 
\end{array}
\right)  
\left(
\begin{array}{l}
u_{j\alpha} \\ v_{j\alpha}
\end{array}
\right) ,
\end{equation}
with $\theta_j\in ]-\pi/4,\pi/4]$ and
\begin{equation}
\tan (2\theta_j) =  \frac{2\chi^0({\bf Q}_j,{\bf
  Q}_j+{\bf K})}{\chi^0({\bf Q}_j+{\bf K},{\bf Q}_j+{\bf K})
  -\chi^0({\bf Q}_j,{\bf Q}_j)} ,
\end{equation}
we obtain the diagonal form
\begin{equation}
F_2 = \sum_{\alpha,j} \bigl( \lambda_j^+ |u_{j\alpha}|^2 +
\lambda_j^- |v_{j\alpha}|^2 \bigr)
\end{equation}
where 
\begin{eqnarray}
\lambda^\pm_j &=&
\frac{g_2}{2}\bigl(2-g_2\chi^0({\bf Q}_j,{\bf Q}_j) 
-g_2 \chi^0({\bf Q}_j+{\bf K},{\bf Q}_j+{\bf K}) \bigr) \pm
\frac{g_2^2}{2} {\rm sgn} \bigl( 
\chi^0({\bf Q}_j+{\bf K},{\bf Q}_j+{\bf K})
-\chi^0({\bf Q}_j,{\bf Q}_j)\bigr)
\nonumber \\ && \times
\bigl( [\chi^0({\bf Q}_j,{\bf Q}_j) 
-\chi^0({\bf Q}_j+{\bf K},{\bf Q}_j+{\bf K})]^2 +4 
[\chi^0({\bf Q}_j,{\bf Q}_j+{\bf K})]^2 \bigr)^{1/2} . 
\label{lambda}
\end{eqnarray}
\eleq
The transition temperature $T_c^j$ is determined by 
\begin{equation}
{\rm min} \, \lambda^{\pm}_j (T_c^j) =0 .
\label{lambda1}
\end{equation}
>From Eqs.~(\ref{lambda}) and (\ref{lambda1}), one easily recovers the
generalized Stoner criterion obtained in the preceding section
[Eq.~(\ref{stoner})]. Since we have assumed $\chi^0({\bf Q}_j,{\bf Q}_j) >
\chi^0({\bf Q}_j+{\bf K},{\bf Q}_j+{\bf K})$ (see
Sec.~\ref{subsec:rpa}), ${\rm min}\,\lambda^{\pm}_j=\lambda^+_j$ and
the order parameter of the transition is $u_{j\alpha}$. For
$T<T_c^j$, we then have $v_{j\alpha}=0$, {\it i.e.} 
\begin{equation}
\Delta_\alpha^{j-}=\tan (\theta_j) \Delta_\alpha^{j+} .
\end{equation}
This equation determines the relative amplitude of the spin modulation
on even and odd chains (see Sec.~\ref{subsec:sm}). 
If $T_c^->T_c^+$ ({\it i.e.} $t_{3b}>0$), then
$u_{-\alpha}\neq 0$ when $T<T_c^-$. To determine whether we can have also
$u_{+\alpha}\neq 0$ at lower temperature ({\it i.e.} coexistence of two
SDW's), we must analyze the quartic contribution to the free energy.

\subsection{Quartic contribution $F_4$} 

\begin{figure}
\epsfysize 7. cm
\epsffile[0 350 300 580]{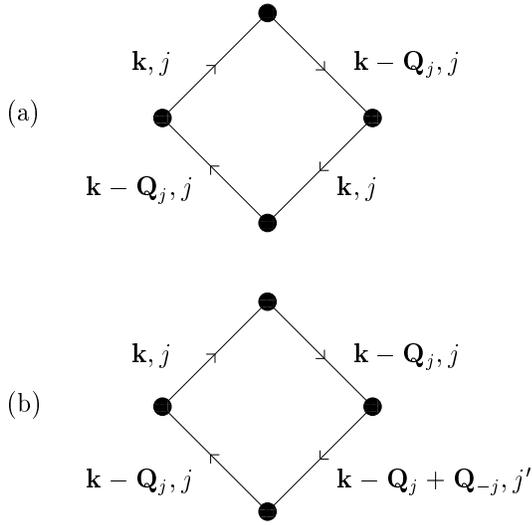}
\caption{
(a) Main contribution to $B_j$. (b) A typical diagram contributing to $C$. 
}
\label{fig:F4}
\end{figure}

The calculation of the quartic contribution $F_4$ is somewhat
lengthy, and we only give the main results. More details can be found in
Appendix \ref{app:I}. Using Eqs.~(\ref{eigenfn}), (\ref{Green}) and
(\ref{op1}), we rewrite the quartic part of the free energy
[Eq.~(\ref{energy1})] as
\begin{eqnarray}
F_4 &=& \sum_\alpha \sum_{j_1\cdots j_4,p_1\cdots p_4} 
\delta_{{\bf Q}_{j_1}+{\bf Q}_{j_3},{\bf Q}_{j_2}+{\bf Q}_{j_4}} 
\nonumber \\ && \times 
\tilde B_\alpha(j_1,j_2,j_3,j_4;p_1,p_2,p_3,p_4) \nonumber \\ && \times 
  {\Delta_\alpha^{j_1p_1}}^* \Delta_\alpha^{j_2p_2}
 {\Delta_\alpha^{j_3p_3}}^* \Delta_\alpha^{j_4p_4} .
\label{F4.0}
\end{eqnarray}
The coefficients $\tilde B_\alpha$ are
defined in Appendix \ref{app:I}. We now express $F_4$ as a function of the
order parameters $u_{j\alpha}$ and $v_{j\alpha}$. Since the formation
of the SDW at wave vector ${\bf Q}_j$ corresponds to the order parameter
$u_{j\alpha}$, we take $v_{j\alpha}=0$ in the following. We then
obtain a free energy of the form
\begin{equation}
F_4 = \sum_\alpha \Bigl\lbrace \sum_j B_j |u_{j\alpha}|^4 + C
|u_{+\alpha} u_{-\alpha}|^2 \Bigr\rbrace .
\label{F4}
\end{equation}
The expression of the coefficients $B_j$ and $C$ is given in
Appendix \ref{app:I}.  The important point here is that the
interaction term in Eq.~(\ref{F4}) is weak, {\it i.e.}
\begin{equation}
\frac{C}{B_j} = O \Biggl( \frac{T^2}{V^2} \Biggr) . 
\end{equation}
This result is easily obtained by considering the diagrams that
contribute to $B_j$ or $C$. The main contribution to $B_j$ comes from
the type of diagrams shown in Fig.~\ref{fig:F4}a. ${\bf Q}_j$ being
the best nesting vector for the pairing in the band $j$, the value of
this diagram is essentially determined by the nesting
properties. Since the quadratic term $F_2$ predicts an instability of
the metallic phase against the formation of a SDW at wave vector ${\bf
Q}_j$, the nesting is good (and becomes better with increasing
$V$). Diagrams contributing to $C$ mix the bands $+$ and $-$. A
typical diagram is shown in Fig.~\ref{fig:F4}b. Contrary to diagrams
of Fig.~\ref{fig:F4}a, it is not possible to have all electronic
states near the Fermi surface and at least one of them has an energy
of order $V$ with respect to the Fermi level. For instance, in the
diagram of Fig.~\ref{fig:F4}b, the state $({\bf k}-{\bf Q}_j+{\bf
Q}_{-j},j')$ has an energy of order $V$ with respect to the Fermi
level when the states $({\bf k},j)$ and $({\bf k}-{\bf Q}_j,j)$ lie
near the Fermi surface. As a result, these diagrams turn out to be of
order $(T/V)^2$ with respect to those contributing to $B_j$ (see
Appendix \ref{app:I}). As shown in the next section, this ensures that
at low temperature the two order parameters $u_{+\alpha}$ and
$u_{-\alpha}$ coexist.

\subsection{Phase diagram}
\label{subsec:pd}

Collecting the results from the preceding sections, we obtain the
following free energy:
\begin{equation}
F=\sum_\alpha \Bigl \lbrace \sum_j \bigl[ A_j |u_{j\alpha}|^2
+ B_j |u_{j\alpha}|^4 \bigr] + C
|u_{+\alpha}u_{-\alpha}|^2 \Bigr\rbrace ,
\label{energy2}
\end{equation}
where $A_j=\lambda^+_j$. 

The free energy $F$ is analyzed in Appendix \ref{app:II}. Assuming
that $T_c^->T_c^+$ ({\it i.e.} $t_{3b}>0$), we find a second-order
metal-SDW transition at temperature 
$T_c^-$, below which $u_{-\alpha}\neq 0$ and $u_{+\alpha}=0$. This
transition opens a gap on the $-$ band, while the $+$ band remains
gapless. This SDW phase is metallic. Note that the minimum of $F$
corresponds to $|u_{j+}|=|u_{j-}|$. This implies that the SDW is
linearly polarized (see Sec.~\ref{subsec:sm}). 

If $4B_+B_--C^2\geq 0$, a second transition takes place at 
\begin{equation}
T_c^{\rm coex} \simeq T_c^+ + \gamma (T_c^+-T_c^-),
\label{Tcoex}
\end{equation}
where $\gamma$ is a constant of order $T^2/V^2$. Below $T_c^{\rm
coex}$, both $u_{-\alpha}$ and $u_{+\alpha}$ are
finite. $u_{+\alpha}$ opens a gap on the $+$ band, making the whole
Fermi surface gapped, so that this
low-temperature SDW is truly insulating. The transition is
also of second order, since the order parameters $u_{-\alpha}$ and
$u_{+\alpha}$ vary continuously at the transition.

Since $C/B_j=O(T^2/V^2)$, the condition $4B_+B_--C^2\geq 0$
is satisfied. Therefore, there is always coexistence of two SDW's at
low temperature when the pairing is intraband. Eq.~(\ref{Tcoex}) shows
that $T_c^{\rm coex}$ is well approximated by $T_c^+$. 

SDW phases with wave vector ${\bf Q}_-$, ${\bf Q}_+$, or ${\bf Q}_-$
and ${\bf Q}_+$, have been previously obtained by Kishigi {\it et al.}
\cite{Kishigi97,Hasegawa98,Miyazaki99,note2} However, the overall phase
diagram (and in particular the existence of a SDW-metal-SDW
transition) as a function of $V$ has not been derived before.

\subsection{Spin modulation $\langle {\bf S}({\bf r})\rangle$}
\label{subsec:sm}

In this section, we determine the spin modulation $\langle {\bf
S}({\bf r})\rangle$ in the different SDW phases of the phase
diagram. ${\bf S}({\bf r})=(1/2)\sum_{\alpha,\sigma,\sigma'}
\hat\psi^\dagger_{\bar\alpha\sigma}({\bf r}) {\bbox\tau}_{\sigma\sigma'}
\hat\psi_{\alpha\sigma'}({\bf r})$ is the spin-density operator and
${\bbox\tau}=(\tau_x,\tau_y,\tau_z)$ stands for the Pauli matrices. 
Noting that $S_-=S_x-iS_y=\sum_\alpha \hat \Delta_{\alpha\uparrow}$, we
obtain 
\begin{eqnarray}
\langle S_-({\bf r})\rangle &=& \sum_{\alpha,j,p} \Delta_\alpha^{jp}
e^{i\alpha({\bf Q}_j+(p-1)\frac{\bf K}{2})\cdot {\bf r}} \nonumber \\
&=& \sum_{\alpha,j} \Delta_\alpha^{j+} e^{i\alpha{\bf Q}_j\cdot {\bf r}}
\bigl[ 1+(-1)^n \tan(\theta_j) \bigr] ,
\label{mod1}
\end{eqnarray}
where we have used $v_{j\alpha}=0$ to obtain the second line of
Eq.~(\ref{mod1}). We consider the more general case where two SDW's
can be present. We write $\Delta_\alpha^{j+}$ as
\begin{equation}
\Delta_\alpha^{j+} =  |\Delta_\alpha^{j+}| e^{i\varphi_\alpha^{j+}} ,
\end{equation}
and introduce the phases $\Theta_j$ and $\phi_j$ defined by
\begin{equation}
\varphi_\alpha^{j+} = \alpha \Theta_j - \phi_j .
\end{equation} 
According to the analysis of the free energy [Sec.~\ref{subsec:pd}
and Appendix \ref{app:II}], 
$|u_{j\alpha}|=|\Delta_\alpha^{j+}/\cos(\theta_j)|$ is independent of
$\alpha$. We therefore obtain 
\begin{eqnarray}
\langle S_-({\bf r})\rangle &=& \sum_j 2|\Delta_+^{j+}| \cos({\bf
  Q}_j \cdot {\bf r}+\Theta_j) e^{-i\phi_j} \nonumber \\ && 
\times \bigl[ 1+(-1)^n \tan (\theta_j) \bigr] .
\label{mod2}
\end{eqnarray}
>From Eq.~(\ref{mod2}), we deduce 
\begin{eqnarray}
\langle S_x({\bf r})\rangle &=& \sum_j 2|\Delta_+^{j+}|
\cos(\phi_j) \cos({\bf Q}_j \cdot {\bf r}+\Theta_j) \nonumber \\ && 
\times \bigl[ 1+(-1)^n \tan (\theta_j) \bigr] , \nonumber \\ 
\langle S_y({\bf r})\rangle &=& \sum_j 2|\Delta_+^{j+}|
\sin(\phi_j) \cos({\bf Q}_j \cdot {\bf r}+\Theta_j) \nonumber \\ && 
\times \bigl[ 1+(-1)^n \tan (\theta_j) \bigr] .
\end{eqnarray}
The phase $\phi_j$ determines the polarization of the SDW's, while
$\Theta_j$ gives their positions with respect to the
underlying crystal lattice. The free energy is independent of $\phi_j$
and $\Theta_j$.

The ratio of the SDW amplitudes on even and odd chains is given by the
factor $\delta_j=(1+\tan(\theta_j))/(1-\tan(\theta_j))$ shown in
Fig.~\ref{fig:theta}. $\delta$ remains close to one when
$V$ is weak ($V\leq V_{c1}$). $\delta \ll 1$ or $\delta \gg 1$  when
$V$ is strong 
($\tan(\theta)\to\pm 1$ for $V\to\infty$) showing that the SDW's become
mostly localized on even or odd chains (depending on the sign of
$\theta$). As expected, $\theta_+$ and $\theta_-$ have opposite
signs so that $\delta_+\simeq \delta^{-1}_-$. 

\begin{figure}
\epsfxsize 7. cm
\epsffile[15 178 275 575]{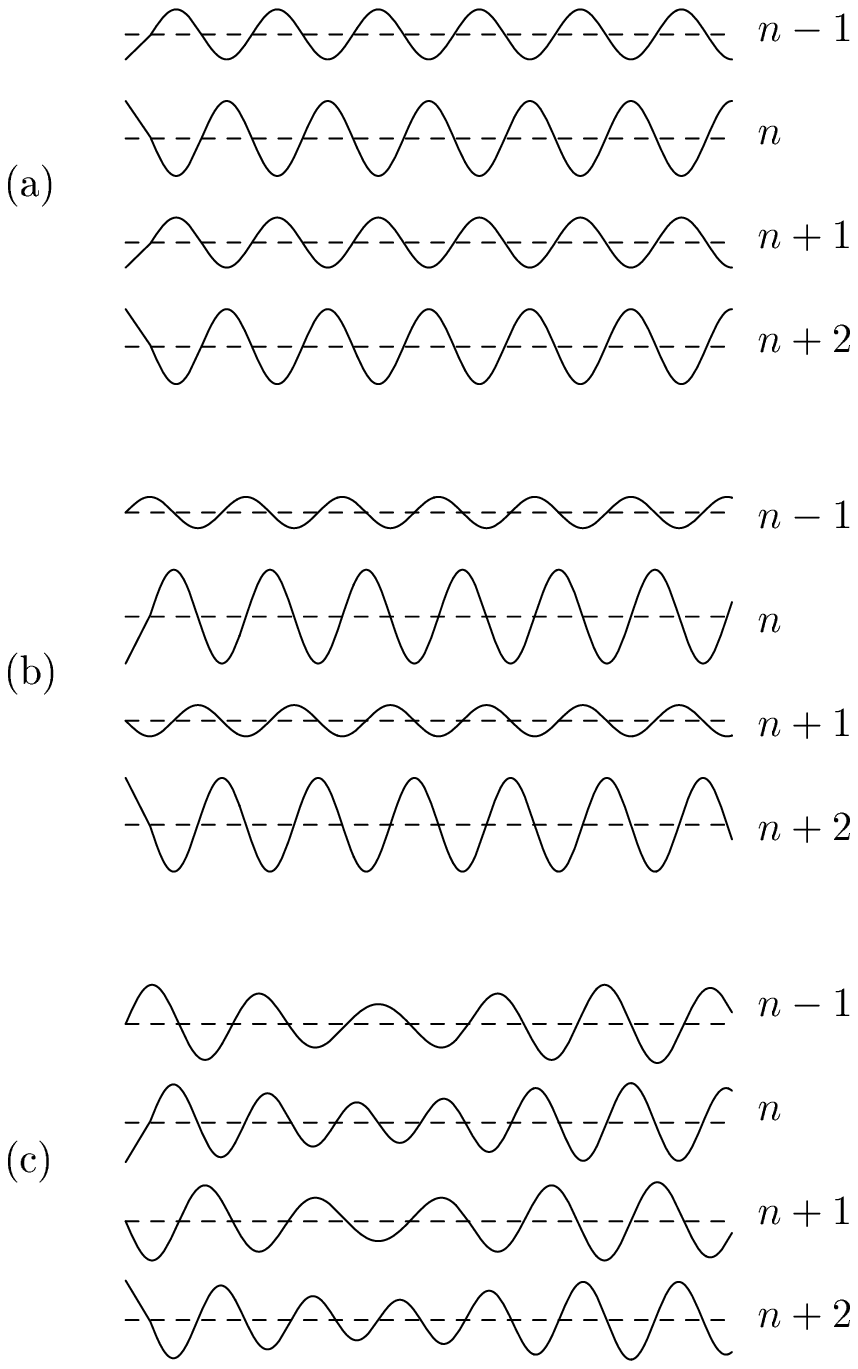}
\caption{
Schematic representation of the spin modulation in the SDW phases ($n$
denotes the chain index). (a) SDW phase for a weak anion 
potential $V$ (interband pairing). (b) SDW phase for a strong anion
potential (intraband pairing). (c) Low-temperature SDW phase
($T<T_c^{\rm coex}$) where two SDW's coexist. }
\label{fig:modulation}
\end{figure}

Consider first the SDW phase occuring for a weak anion potential $V$ 
(interband pairing). There is a single SDW with a wave vector ${\bf
Q}=(2k_F,\pi/b)$. The spin 
modulations on two neighboring chains are out-of-phase and the
ratio $\delta$ of their amplitudes is close
to one (Fig.~\ref{fig:modulation}a).

The SDW's corresponding to intraband pairing are also commensurate in
the transverse direction, but with $(Q_y)_j=\pi/2b$: the spin
modulations on two neighboring chains are in phase
quadrature. Fig.~\ref{fig:modulation}b shows the spin modulation in the 
phase with a single SDW at wave vector ${\bf Q}_-$. The spin
modulations below $T_c^{\rm coex}$ when two SDW's coexist is shown in
Fig.~\ref{fig:modulation}c. The beating phenomenon is due to the fact
that $(Q_x)_-\neq (Q_x)_+$. 

\section{Conclusion}

Anion ordering in the organic conductor (TMTSF)$_2$ClO$_4$ is expected
to strongly influence the SDW instability. While a weak anion potential
($V\lesssim t_b$) 
suppresses the SDW instability, a strong anion potential leads to a
rich phase diagram. When $V$ becomes of the order of the interchain
hopping amplitude
($V\sim t_b$), the effective hopping between even and odd chains is
reduced. This opens up the possibility to have two successive
instabilities when the temperature decreases. The first one primarily
occurs on even (or odd) chains and destroys the Fermi surface of
one of the two electronic bands. The other electronic band remains
metallic in this phase.  At lower temperature, a
second transition occurs, primarily on odd (or even) chains, making
the whole Fermi surface gapped.  

According to Ref.~\onlinecite{Peleven01}, the gap due to the anion
potential in the electronic dispersion is of the order of the
transverse bandwidth, and the intraband nesting is almost perfect. In
our model, this corresponds to a large value of $V$, {\it i.e.}
$V\gtrsim t_b$. In this situation, one expects the ground-state to be
a SDW. Experimentally, the ground-state is found to be a superconductor.
There is no contradiction with our model though, since we did
not consider the possibility of a superconducting transition. However,
the observation of FISDW phases in (TMTSF)$_2$ClO$_4$ does require
deviations from perfect nesting (otherwise, only the phase
$N=0$ would be observed). This suggests that even when the anion
potential is strong (large $V$ in our model), important deviations
form perfect nesting persist. The latter can be taken into account by
considering $t_{4b}$ in our model. Alternatively, the
observation of the FISDW phases could indicate that the anion
potential is weaker than what is predicted by quantum chemistry
calculation. In our model, this would correspond to the region
$V_{c1}<V<V_{c2}$, where the ground-state is metallic. [Note that
$V_{c1}=0$ if $t_{2b}$ is large enough.] 

Two successive transitions have been observed in (TMTSF)$_2$ClO$_4$ at
high magnetic field.\cite{McKernan95,Moser99,Chung00} The possibility
that each of these transitions corresponds to an intraband pairing,
thus partially destroying the Fermi surface, has been suggested early
on. \cite{McKernan95} Although our
conclusions do not apply {\it stricto sensus} to the experimental
situation, since they are restricted to the zero-field case, they
indicate that this  scenario could indeed take
place in (TMTSF)$_2$ClO$_4$ at high field, in agreement with the
conclusions of Refs.~\cite{Kishigi97,Hasegawa98,Miyazaki99,note2}

Beside the existence of two successive transitions at low temperature
when $V$ is strong enough, an important result obtained in this paper
is the overall phase diagram as a function of $V$. In particular, we
have obtained a SDW-metal-SDW transition at low temperature
(Figs.~\ref{fig5} and \ref{fig6}).  

Finally, we note that our results seriously call into question the
validity of most of the previous works on the SDW transition at high
magnetic field in (TMTSF)$_2$ClO$_4$. In the presence of anion
ordering, the SDW order parameter necessarily has two Fourier
components, and the standard Stoner criterion cannot be used
anymore. Instead, one should consider the generalized Stoner criterion
obtained in Sec.~\ref{subsec:rpa} [Eq.~(\ref{stoner})]. 

{\it Note added:} After completion of this work, we became aware of a
related work by Zanchi and Bjeli\v{s}.\cite{Zanchi01} These authors
have considered the effect of anion ordering on the SDW instability in
(TMTSF)$_2$ClO$_4$ and obtained the transition temperature
within RPA. Their results are similar to ours. However, they did not
take into account a finite $t_{3b}$ and did not study the
coexistence of the two SDW's with wave vectors ${\bf Q}_+$ and ${\bf
Q}_-$ at low temperature. 

\section*{Acknowledgment}

We thank D. Zanchi for a useful comment on
Ref.~\onlinecite{Zanchi01}. One of the authors (KS) would like to
thank Victor M. Yakovenko for support during the work.

\bleq

\appendix

\section{Quartic contribution $F_4$ to the free energy}
\label{app:I}

Using Eqs.~(\ref{eigenfn}), (\ref{Green}) and
(\ref{op1}), the quartic part of the free energy
[Eq.~(\ref{energy1})] can be written as in Eq.~(\ref{F4.0}) with 
\begin{eqnarray}
\tilde B_\alpha(j_1,j_2,j_3,j_4;p_1,p_2,p_3,p_4) &=& \frac{g_2^4}{2N_\perp}
\sum_{k_y} \sum_{j_5\cdots j_8} \sum_{p_5\cdots p_8} \sum_{p_5'\cdots p_8'}
F_\alpha(j_1\cdots j_8,p_1\cdots p_8,p_5'\cdots p_8';k_y)
K_\alpha(j_1\cdots j_8;k_y) \nonumber \\ && \times
\sum_{n=-\infty}^\infty
\delta_{\alpha(p_1-p_2+p_3+p_4)\frac{K}{2}
  +(-p_5-p_6-p_7-p_8+p_5'+p_6'+p_7'+p_8')\frac{K}{2}, n\frac{2\pi}{b}} ,
\\ 
K_\alpha(j_1\cdots j_8;k_y) &=& \frac{T}{bL_x} \sum_{k_x,\omega}
(i\omega-\epsilon^{j_5}_{\alpha,{\bf k}})^{-1} 
(i\omega-\epsilon^{j_6}_{\bar\alpha,{\bf k}-\alpha{\bf Q}_{j_2}})^{-1} 
 \nonumber \\ && \times
(i\omega-\epsilon^{j_7}_{\alpha,{\bf k}-\alpha{\bf
    Q}_{j_2}+\alpha{\bf Q}_{j_3}})^{-1} 
(i\omega-\epsilon^{j_8}_{\bar\alpha,{\bf k}-\alpha{\bf Q}_{j_1}})^{-1}, 
\nonumber \\ 
F_\alpha(j_1\cdots j_8,p_1\cdots p_8,p_5'\cdots p_8';k_y) &=& 
\gamma_{k_y}^{j_5p_5} \gamma_{k_y}^{j_5p_5'}
\gamma_{k_y-\alpha Q_{j_2y}-(\alpha(p_2-1)-p_5'+p_6)\frac{K}{2}}^{j_6p_6}
\gamma_{k_y-\alpha
  Q_{j_2y}-(\alpha(p_2-1)-p_5'+p_6)\frac{K}{2}}^{j_6p_6'}
\nonumber \\ && \times 
\gamma_{k_y-\alpha Q_{j_2y}+\alpha Q_{j_3y}
  -(\alpha(p_2+p_3)-p_5'+p_6-p_6'+p_7)\frac{K}{2}}^{j_7p_7}
\nonumber \\ && \times 
\gamma_{k_y-\alpha Q_{j_2y}+\alpha Q_{j_3y} 
  -(\alpha(p_2+p_3)-p_5'+p_6-p_6'+p_7)\frac{K}{2}}^{j_7p_7'} 
\nonumber \\ && \times 
\gamma_{k_y-\alpha Q_{j_1y}-(\alpha(p_1-1)-p_5+p_8')\frac{K}{2}}^{j_8p_8} 
\gamma_{k_y-\alpha
  Q_{j_1y}-(\alpha(p_1-1)-p_5+p_8')\frac{K}{2}}^{j_8p_8'} ,
\end{eqnarray}
where $N_\perp=L_y/b$ is the total number of chains.
Since $v_{j\alpha}=0$, we have 
\begin{eqnarray}
\Delta_\alpha^{jp} &=& c_{jp} u_{j\alpha} , \nonumber \\
c_{j+} &=& \cos(\theta_j) , \,\,\,\, c_{j-} = \sin(\theta_j) . 
\end{eqnarray}
Noting that the condition ${\bf Q}_{j_1}+{\bf Q}_{j_3}={\bf
Q}_{j_2}+{\bf Q}_{j_4}$ implies $j_1=j_2=j_3=j_4$ or
$j_1=j_2=-j_3=-j_4$ or $j_1=-j_2=-j_3=j_4$, we obtain Eq.~(\ref{F4})
with
\begin{eqnarray}
B_j &=& \sum_{p_1\cdots p_4} c_{jp_1}c_{jp_2}c_{jp_3}c_{jp_4} 
\tilde B_\alpha(j,j,j,j;p_1,p_2,p_3,p_4) , \nonumber \\ 
C &=& \sum_j \sum_{p_1\cdots p_4} \bigl[ c_{jp_1}c_{jp_2}c_{-jp_3}c_{-jp_4}
\tilde B_\alpha(j,j,-j,-j;p_1,p_2,p_3,p_4) 
\nonumber \\ && 
+  c_{jp_1}c_{-jp_2}c_{-jp_3}c_{jp_4}
\tilde B_\alpha(j,-j,-j,j;p_1,p_2,p_3,p_4) \bigr] . 
\end{eqnarray}

The main contribution to $B_j$ comes from the type of diagrams shown in
Fig.~\ref{fig:F4}. This diagram involves the quantity
\begin{equation}
K_\alpha(j_1=\cdots=j_8=i;k_y) = \frac{T}{bL_x} \sum_{k_x,\omega} 
(i\omega-\epsilon^j_{\alpha,{\bf
    k}})^{-2}(i\omega-\epsilon^j_{\bar\alpha,{\bf k}-\alpha{\bf
    Q}_j})^{-2} .
\end{equation}
To evaluate the order of magnitude of this diagram, we can assume
perfect nesting. We then find
\begin{equation}
K_\alpha(j_1=\cdots=j_8=i;k_y) = \frac{T}{bL_x} \sum_{k_x,\omega} 
\Bigl(\omega^2+{\epsilon^j_{\alpha,{\bf k}}}^2\Bigr)^{-2} \sim
\frac{1}{T^2} .  
\end{equation}

A typical contribution to $C$ is shown in Fig.~\ref{fig:F4}. The
order of magnitude of the diagram is given by
\begin{equation}
K_\alpha(j,j,-j,-j,j,j,j_7,j;k_y) = \frac{T}{bL_x} \sum_{k_x,\omega} 
(i\omega-\epsilon^j_{\alpha,{\bf k}})^{-1}
(i\omega-\epsilon^j_{\bar\alpha,{\bf k}-\alpha{\bf Q}_j})^{-2}
(i\omega-\epsilon^{j_7}_{\alpha,{\bf k}-\alpha{\bf Q}_j+\alpha{\bf
    Q}_{-j}})^{-1} .
\end{equation}
To evaluate the preceding equation, we can assume perfect nesting in
the $j$ band:
\begin{equation}
K_\alpha(j,j,-j,-j,j,j,j_7,j;k_y) = \frac{T}{bL_x} \sum_{k_x,\omega} 
(i\omega-\epsilon^j_{\alpha,{\bf k}})^{-1}
(i\omega+\epsilon^j_{\alpha,{\bf k}})^{-2}
(i\omega-\epsilon^j_{\alpha,{\bf k}}+a)^{-1}
\sim \frac{1}{V^2} ,
\end{equation}
\eleq
where $|a|=|\epsilon^j_{\alpha,{\bf k}}-\epsilon^{j_7}_{\alpha,{\bf
k}-\alpha{\bf Q}_j+\alpha{\bf Q}_{-j}}| \sim V$. We therefore conclude
that $C/B_j=O(T^2/V^2)$.

\section{Minimum of the free energy $F$}
\label{app:II} 

The minimum of the free energy is obtained by solving the equations
$\partial F/\partial u_{j\alpha}^*=0$: 
\begin{eqnarray}
u_{+\alpha} [A_++2B_+|u_{+\alpha}|^2
+C|u_{-\alpha}|^2] &=& 0 ,  \nonumber \\ 
u_{-\alpha} [A_-+2B_-|u_{-\alpha}|^2
+C|u_{+\alpha}|^2] &=& 0 . 
\label{minF}
\end{eqnarray}

In the metallic phase, $u_{j\alpha}=0$ and $F=0$. Below $T_c^-$, there
is a phase with $u_{-\alpha}\neq 0$
and $u_{+\alpha}=0$. From Eqs.~(\ref{minF}) and (\ref{energy2}), we deduce
\begin{equation}
|u_{-\alpha}|^2 = - \frac{A_-}{2B_-} , \,\,\,\,\, 
F_- = - \frac{A_-^2}{2B_-} .
\end{equation}
Let us now consider a phase with two coexisting order parameters:
$u_{+\alpha},u_{-\alpha} \neq 0$. Eqs.~(\ref{minF}) and (\ref{energy2})
yield 
\begin{eqnarray}
|u_{+\alpha}|^2 &=& \frac{-2A_+B_-+A_-C}{4B_+B_--C^2} , \nonumber \\ 
|u_{-\alpha}|^2 &=& \frac{-2A_-B_++A_+C}{4B_+B_--C^2} , \nonumber \\ 
F_{\rm coex} &=& 2\frac{-A_+^2B_- - A_-^2B_+ + A_+A_-C}{4B_+B_--C^2} .
\label{Fcoex} 
\end{eqnarray}
This solution is allowed only if $|u_{j\alpha}|\geq 0$. The phase
with two coexisting order parameters will be observed only if it has a
lower free energy than the phase with a single order parameter,
{\it i.e.} $F_{\rm coex} \leq F_-$:
\begin{equation}
2\frac{-A_+^2B_- - A_-^2B_+ + A_+A_-C}{4B_+B_--C^2} \leq -
\frac{A_-^2}{2B_-}. 
\end{equation}
For $4B_+B_--C^2\leq 0$, this condition can be rewritten as
$-(2A_-B_+-A_+C)^2 \geq 0$, which shows that coexistence is not
possible. For $4B_+B_--C^2\geq 0$, the condition $F_{\rm coex} \leq
F_-$ becomes $-(2A_-B_+-A_+C)^2 \leq 0$ and is therefore always
satisfied. We thus conclude that coexistence occurs when the condition 
\begin{equation}
4B_+B_--C^2\geq 0
\end{equation}
is fulfilled. The corresponding transition temperature $T_c^{\rm
coex}$ is then determined from $|u_{j\alpha}|\geq 0$
[Eqs.~(\ref{Fcoex})]. When $C/B_j\ll 1$, $T_c^{\rm coex}$ is obtained from
$|u_{+\alpha}|\geq 0$. Writing $A_j=a_j(T-T_c^j)$ and neglecting the
temperature dependence of $B_j$ and $C$, we obtain
\begin{eqnarray}
T_c^{\rm coex} &=& T_c^+ + \frac{a_-C}{2a_+B_-}(T_c^{\rm coex}-T_c^-)
\nonumber \\ 
& = &  T_c^+ + \gamma (T_c^+-T_c^-) + O(T^4/V^4) ,
\end{eqnarray}
where $\gamma = a_-C/(2a_+B_-)$ is a constant of order $O(T^2/V^2)$.

\ecols



\begin{thebibliography}{99}
\bibitem{note2} It should be pointed out that the work reported in 
  Refs.~\onlinecite{Kishigi97,Hasegawa98,Miyazaki99} assumes a SDW
  order parameter with a single Fourier component, whereas there are
  (at least) two Fourier components in the presence of the anion
  potential (see text).
\end{thebibliography}
\end{document}